\documentclass[letterpaper,twocolumn,amsmath,amssymb]{revtex4-2}
\usepackage{siunitx}
\usepackage{graphicx}
\usepackage{here}
\usepackage{booktabs}
\usepackage{setspace}
\usepackage[svgnames]{xcolor}
\usepackage[unicode,colorlinks=true,citecolor=magenta,linkcolor=magenta,urlcolor=magenta]{hyperref}
\setcitestyle{numbers,comma,sort&compress}
\usepackage{url}
\usepackage[normalem]{ulem}
\usepackage[nameinlink,capitalize]{cleveref}
\usepackage{mhchem}
\usepackage{enumerate}

\usepackage{hhline}
\usepackage{makecell}
\newcommand{\etal}{\textit{et~al.}}
\begin{document}
\title{On the Author Correction to ``Magnetic flux trapping in hydrogen-rich high-temperature superconductors'', Nat Phys. 19, 1293 (2023)}
\author{N. Zen}
\affiliation{National Institute of Advanced Industrial Science and Technology,\\Tsukuba Central 2-12, Ibaraki 305-8568, Japan}
\begin{abstract}
  In Fig.~4c, under the section titled ``Pinning and thermally activated motion of vortices'' in \href{https://doi.org/10.1038/s41567-023-02089-1}{Nat Phys. 19, 1293 (2023)}~\cite{natp}, Minkov and co-workers presented the time dependence of the magnetic moment of sulfur hydride (\ce{H3S}) under high pressure and argued that they had observed magnetic flux creep at 165~K, 180~K and 185~K. Flux creep is a phenomenon observed under the assumption that the material under study can trap magnetic flux, and thus, Fig.~4c serves as evidence that \ce{H3S} traps magnetic flux and is a high-temperature superconductor. The claim remains unchanged even in the recently published \href{https://doi.org/10.1038/s41567-025-02823-x}{Author Correction}~\cite{natpac} to Ref.~\cite{natp}. However, Ref.~\cite{natpac} discloses an experimental protocol they used to collect the time-dependent magnetic moment data. In this Commentary Paper, we point out that the protocol is not applicable to \ce{H3S} under high pressure and propose an alternative protocol. The correct protocol demonstrates that the claim in Refs.~\cite{natp,natpac}---that their time-dependent magnetic moment data serve as evidence of ``pinning and thermally activated motion of vortices''---is indeed invalid.
\end{abstract}
\maketitle
Unlike the original version of Ref.~\cite{natp}, Minkov~\etal~\cite{natpac} recognized the importance of $t_{delay}$, the time interval between turning off the external magnetic field and starting the measurement of the magnetic moment. If there is thermally activated motion of vortices in a material, the magnetic moment will change logarithmically with time~\cite{Kim1962,AndKim1964}. The longer the delay in $t_{delay}$, the longer the measurement period required. (The reason for this has already been explained elsewhere~\cite{Zenphysc}.) Therefore, for phenomena with logarithmic time dependence such as flux creep, it may be useful to introduce a measurement period index defined by the following equation:
\begin{equation}
p\equiv \log\frac{t_{end}}{t_{delay}},\label{eq:p}
\end{equation}
where $t_{end}$ denotes the end time of the measurement. A larger value of $p$ indicates that phenomena have been observed over a longer duration. However, in practice, it is not self-evident what range of $p$ values is appropriate for flux creep. In this study, we estimate a suitable range of the $p$-value by referring to the description in \href{https://doi.org/10.1038/s41567-025-02823-x}{Author Correction}~\cite{natpac}, and on this basis re-evaluate the experimental results reported therein.

Ref.~\cite{natpac}, the Author Correction to Ref.~\cite{natp}, states, ``\emph{The $t_{delay}$ can vary widely between different flux creep experiments\textsuperscript{1}, ranging from hundreds of microseconds\textsuperscript{2} to several hours\textsuperscript{3}.}'' Here, ref.\textsuperscript{3}---Guven~\etal~(2024)~\cite{Guven2024}---is the protocol that Minkov and co-workers claim to have followed, as the description then continues: ``\emph{Our experiments followed a protocol used in applied superconductivity\textsuperscript{3}, where the delay time in long-term flux creep studies is several hours (for example, $t_{delay}=5~h$, see ref.\textsuperscript{3}), and the magnetization field is $\si{\mu}_{0}H_{M}=0.5-1.0~T$ (see refs.\textsuperscript{3,6}). Therefore, in the article, we magnetized the \ce{H3S} at $T=165~K$ with an applied magnetic field $\si{\mu}_{0}H_{M}=1~T$, and implemented a delay time of $t_{delay}=38~h$.}'' Their statement gives the impression that their experiments were conducted in accordance with ref.\textsuperscript{3}, as if that protocol had already been available at the time. However, ref.\textsuperscript{3} was published on December 5, 2023, more than a year after Minkov and co-workers submitted their paper to {\it Nature Physics} on September 21, 2022. In other words, they refer to a protocol that did not exist at the time of their experiments. This means that \href{https://doi.org/10.1038/s41567-025-02823-x}{Author Correction}~\cite{natpac} issued by {\it Nature Physics} contains a justification that is inconsistent with the facts, yet it has been left unaddressed for a year. However, correcting that point is not the aim of this paper. Rather, this work accepts this seemingly constructed narrative for the sake of argument and attempts to demonstrate that ref.\textsuperscript{3}---Guven~\etal~(2024)~\cite{Guven2024}---is not an appropriate experimental protocol for their flux creep measurements. The values of $t_{delay}$ and $t_{end}$ used in ref.\textsuperscript{3}, Guven~\etal~(2024)~\cite{Guven2024}, are 5~$h$ and 30~$h$, respectively (see Fig.~8 in Ref.~\cite{Guven2024}); accordingly, the corresponding $p$-value is approximately 0.78 using \cref{eq:p}.

\begin{center}
\begin{figure*}[h!b]
\centering
\includegraphics[width=0.9\textwidth]{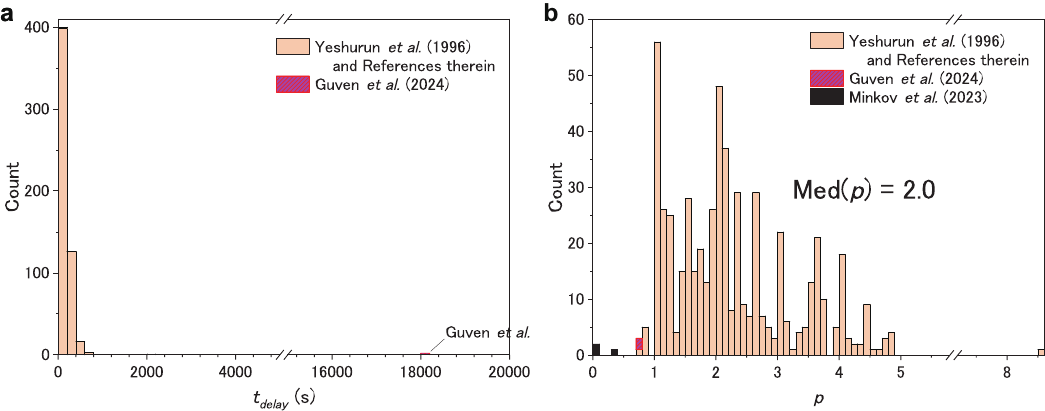}
\caption{{\bf a}, Distribution of $t_{delay}$ compiled from 546 experimental flux creep data reported in Yeshurun~\etal~(1996)~\cite{Yeshurun1996} and the references therein, which is the review article cited by Minkov~\etal~\cite{natpac}. Guven~\etal~(2024)~\cite{Guven2024} is the protocol that Minkov and co-workers claim to have followed. {\bf b}, Distribution of $p$ compiled from these articles.}
\label{fig1}
\end{figure*}
\end{center}

First, we point out that there is no valid reason to adopt a specific $p$-value in a particular paper as the protocol. As Minkov and co-workers themselves state in their \href{https://doi.org/10.1038/s41567-025-02823-x}{Author Correction}~\cite{natpac}, ``\emph{$t_{delay}$ can vary widely between different flux creep experiments\textsuperscript{1}}''; that is, $t_{delay}$ has a distribution, and accordingly, the $p$-value does as well. Using the review article ref.\textsuperscript{1} in \href{https://doi.org/10.1038/s41567-025-02823-x}{Author Correction}~\cite{natpac}---Yeshurun~\etal~(1996)~\cite{Yeshurun1996}---along with all the experimental data cited therein, the distribution of $t_{delay}$ is shown in \cref{fig1}a, and the corresponding distribution of $p$-values is shown in \cref{fig1}b. The median of $p$ is 2.0. (Details of these data processing procedures are given in the Methods section.) On the other hand, the $p$-values used in Minkov~\etal~are 0.38 for the 165~K data and less than 0.1 for the 180~K and 185~K data, which are considerably smaller than the median value of 2.0, and even smaller than Guven~\etal's $p=0.78$~\cite{Guven2024}, which they claim to be a legitimate protocol.
\onecolumngrid
\twocolumngrid

Second, the time dependence of the trapped field data in Guven~\etal~(2024)~\cite{Guven2024} is intended for extracting the joint resistance of \ce{MgB2} coils, not for demonstrating thermally activated motion of vortices (TAMV) in \ce{MgB2}. Since the time dependence of the magnetic moment data in Minkov~\etal~(2023)~\cite{natp} is aimed to demonstrate TAMV in \ce{H3S} under high pressure, Guven~\etal's $p$-value should not be adopted, let alone an even smaller value.

The discovery of superconductivity in \ce{MgB2}~\cite{Naga2001} was accompanied by the demonstration of the Meissner effect, and shortly thereafter, magnetic hysteresis loops---evidence that persistent currents circulate within the material---were reported by other groups~\cite{Finnemore2001,nat2001}. Thus, the magnetic evidence confirming \ce{MgB2} as a superconductor is well established. In contrast, magnetic evidence establishing superconductivity in \ce{H3S} has not yet been presented. The 2015 publication by the same group~\cite{nat2015}, widely regarded as a breakthrough in superconductivity, presented a difference in temperature-dependent ZFC and FC magnetization as evidence of the Meissner effect, but, of course, this does not constitute an observation of the Meissner effect. The Meissner effect can only be claimed when the expulsion of magnetic fields from the interior of the material---such as a decrease in magnetization in FC measurements---is observed. Since \ce{H3S} under high pressure does not show a decrease in magnetization in FC measurements (see Fig.~4a in Ref.~\cite{nat2015}), the Meissner effect cannot be acknowledged. In addition, the same article includes experimental data resembling magnetic hysteresis loops (Fig.~4c in Ref.~\cite{nat2015}), but the data are also open to question, as no group has been able to reproduce them even after more than a decade. More seriously, signs of inexplicable data manipulation have been found~\cite{HKphysc2025} in supposedly magnetic hysteresis data published by the same group in 2022~\cite{natc} and 2023~\cite{natcac}. On this point, an Editor's Note was published on March 6, 2024, but the issue remains unresolved even two years later. Influential scientists conclude that ``\emph{hydride superconductivity is real}''~\cite{heretostay}, whereas others contend that this conclusion is premature~\cite{Gabovich2025}. Although no data manipulation can be tolerated, this matter provides a valuable opportunity to renew our recognition that it is always up to us to determine what is true.

Finally, it should be pointed out that a recent theoretical study indicates that the superconducting transition temperature of metallic hydrogen does not exceed 30~K~\cite{Marel2025}. This suggests that the data on hydride superconductivity should be examined more carefully and with greater objectivity, contrary to the prevailing view based on Ashcroft's optimistic calculations~\cite{Ash1968,Ash2004}.

Since \ce{MgB2} has sufficient magnetic evidence, $p=0.78$ used in Guven~\etal~(2024)~\cite{Guven2024} is not an issue.{\if0 As previously mentioned, Ref.~\cite{Guven2024} is not intended to prove TAMV in the first place.\fi}{\if0 Similarly, the time-dependent magnetic data of \ce{Ba_{0.6}K_{0.4}Fe2As2}~\cite{Yamamotoyama} referenced by Minkov~\etal~(ref.\textsuperscript{7} in Ref.~\cite{natpac} or ref.\textsuperscript{15} in their Reply to this author~\cite{AR2Z}) pose no issues, since its magnetic evidence for superconductivity is also sufficient~\cite{Rotter2008,Kim2009}.\fi} In contrast, \ce{H3S} under high pressure lacks decisive magnetic evidence to support its claimed superconductivity. Since it has not shown the Meissner effect or magnetic hysteresis loops, a sufficiently large $p$-value must be used if one wishes to claim the occurrence of thermally activated motion of vortices (TAMV) in the material.

\begin{center}
\begin{figure*}[h!b]
\centering
\includegraphics[width=0.9\textwidth]{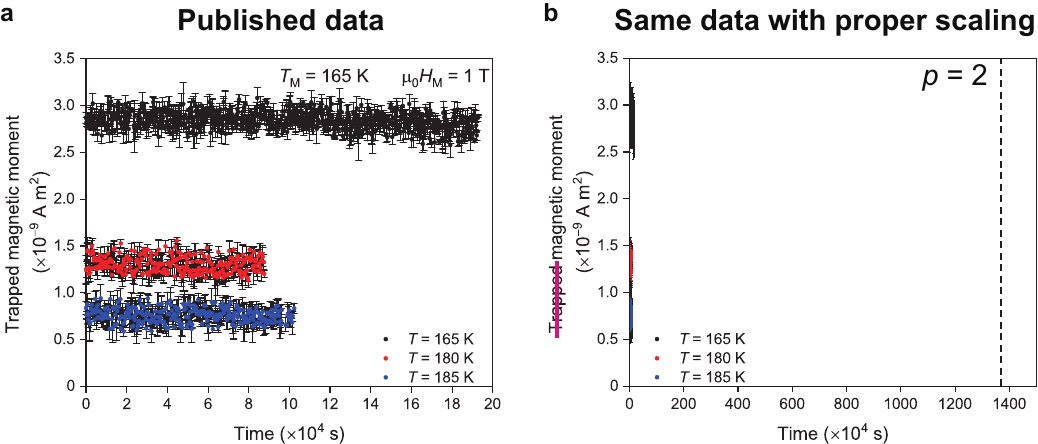}
\caption{{\bf a}, Fig.~4c of Refs.~\cite{natp,natpac}, and {\bf b}, that with proper scaling. The vertical dashed line represents the $p=2$ boundary for the $T=165$~K data with $t_{delay}=38$~h. According to \cref{eq:p}, it is at $1368\times 10^{4}$~s. The $p=2$ lines for the $T=180$~K data with $t_{delay}=95$~h and the $T=185$~K data with $t_{delay}=133$~h are at $3420\times 10^{4}$~s and $4788\times 10^{4}$~s, respectively, and are out of frame.}
\label{fig2}
\end{figure*}
\end{center}

None of the materials used in Yeshurun~\etal~(1996)~\cite{Yeshurun1996} or in any of the references therein were claimed to be superconducting based solely on time-dependent magnetic moment data. (Indeed, most of them were either YBCO or BSCCO.) However, for the sake of argument, we will consider $p=2$, as extracted from \cref{fig1}b, to be a sufficiently large value to claim TAMV. Then, Minkov~\etal's data, properly plotted on the appropriate scale, would look like \cref{fig2}b. As seen, there is no data sufficient to claim the occurrence of TAMV. Since there is no data for TAMV, it is obviously not possible to fit the data to determine the flux creep rate $S$. In \href{https://doi.org/10.1038/s41567-025-02823-x}{Author Correction}~\cite{natpac}, they gave up showing $S$ and instead estimated the decay rate as a percentage, but it amounts the same thing. Flux creep must be assessed on the scale of \cref{fig2}b. From visual inspection of \cref{fig2}b alone, it is impossible to determine whether the magnetic moment will decay at later times or be essentially decay-free; it could even show an increasing trend. One should not rule out such a possibility by simply asserting that such a phenomenon would be absurd. Science must give priority above all else to measured facts, and it should be built upon them. Only after it has been firmly established that the magnetic moment indeed decays should one discuss a decay rate, whether expressed as $S$ or as a percentage.

Thus, the claim by Minkov~\etal~\cite{natp,natpac} that $m_{trap}$ in \ce{H3S} is practically decay-free is invalid. Figure 4c of Refs.~\cite{natp,natpac} (reproduced here as \cref{fig2}a) presents only a portion of the time-dependent data that should have been obtained; no matter how much a single scene of a film is stretched, it cannot represent the whole. Comparing such essentially blank data with the practically decay-free data of established superconductors such as \ce{MgB2}~\cite{Guven2024,Naga2001,Finnemore2001,nat2001} and \ce{Ba_{0.6}K_{0.4}Fe2As2}~\cite{Yamamotoyama,Rotter2008,Kim2009}, as done in Refs.~\cite{natp,natpac}, is scientifically meaningless.
{}

Following the previously claimed Meissner effect~\cite{nat2015} and diamagnetic hysteresis loops~\cite{nat2015,natc,natcac}, the present flux creep assuming flux trapping~\cite{natp,natpac} also fails to constitute experimental evidence to support ``\emph{hydride superconductivity}~\cite{heretostay}''. It is clear from Refs.~\cite{Hjsuper2024,Sridhar2023} that electrical zero resistance alone cannot constitute proof of superconductivity. Therefore, no matter how many electrical data suggestive of zero resistance, such as those reported in Refs.~\cite{Eremets2007-1126,Hemley2015-0417,Eremets2016,Troyan2022-0512,ShimizuEremets2015-0909,natcommun2019-0212,Shimizu2019-0926,Shimizu2019-1001,Eremets2020-0516,Som2018-0823,Shimizu2018-0907,Dro2018-1204,sciadv2019-0413,mattod2019-0811,Cui2019-1002,Cui2019-1208,natcommun2020-0612,Hong2020-0916,advmat2020-1008,Sun2020-1027,Dias2020-1117,mattod2020-1227,Chen2021-0109,Kong2021-0323,Wuhao2021-0629,Hong2021-0710,Cui2021-0712,Ma2021-1011,Li2021-1130,frontiers2021-1216,CQJ2021-1231,Friedemann2022-0321,Wuhao2022-0328,advmat2022-0504,YMM2022-0510,Bi2022-0513,Wang2022-0515,Shimizu2022-0602,CQJ2022-0701,Chen2022-0826,Cui2022-1202,CQJ2023-0316,CQJ2023-0320,Pudalov2023-0603,Pudalov2023-0608,Struzhkin2023-0808,Friedemann2023-0817,CQJ2023-0915pub,CQJ2023-0928,Cui2023-1031,Cui2023-1209,Song2023-1215,Cui2023-1220}, are accumulated{\if0 (even if those data were obtained by appropriate methods, processed without raising any concerns, and made publicly available in raw form)\fi}, the ground state of ``\emph{hydride superconductivity}'' cannot be regarded as superconducting. An inhomogeneous granular mixture of semiconducting and metallic phases is the most plausible scenario, as discussed in Refs.~\cite{Hjsuper2024,Ren2025}.
\onecolumngrid
\twocolumngrid
\section*{Methods}\label{sec:meth}
The statistics in \cref{fig1} are based on 83 experimental papers referenced in Yeshurun~\etal~(1996)~\cite{Yeshurun1996}, with a total of 546 data. Yeshurun~\etal~(1996)~\cite{Yeshurun1996} is the review article referenced in \href{https://doi.org/10.1038/s41567-025-02823-x}{Author Correction}~\cite{natpac}. The values for $t_{delay}$ and $t_{end}$ were used as mentioned in the paper, if provided. If not mentioned, they were extracted from the graph. To ensure fairness, the values were read from the plots, not from the extrapolated lines. Magnetic relaxation data in a field were also included. The $t_{delay}$ here refers to the interval between the time when the magnetic field reaches the specified value and when the measurement starts. On the other hand, data with the vertical axis representing electrical characteristics such as current density were not included in the statistics, as only magnetic properties are of interest. Additionally, data where the magnetic values on the vertical axis were merely converted to their derivatives or potential energy were excluded, to avoid duplication. If the reference list in Yeshurun~\etal~(1996)~\cite{Yeshurun1996} is numbered consecutively from the top, the 83 papers used for the statistics are 2~\cite{2}, 9~\cite{9}, 12~\cite{12}, 18~\cite{18}, 22~\cite{22}, 30~\cite{30}, 42~\cite{42}, 45~\cite{45}, 72~\cite{72}, 73~\cite{73}, 76~\cite{76}, 79~\cite{79}, 82~\cite{82}, 83~\cite{83}, 89~\cite{89}, 96~\cite{96}, 101~\cite{101}, 102~\cite{102}, 109~\cite{109}, 117~\cite{117}, 121~\cite{121}, 122~\cite{122}, 126~\cite{126}, 127~\cite{127}, 135~\cite{135}, 137~\cite{137}, 139~\cite{139}, 140~\cite{140}, 141~\cite{141}, 142~\cite{142}, 147~\cite{147}, 150~\cite{150}, 151~\cite{151}, 153~\cite{153}, 154~\cite{154}, 157~\cite{157}, 159~\cite{159}, 162~\cite{162}, 174~\cite{174}, 175~\cite{175}, 176~\cite{176}, 178~\cite{178}, 184~\cite{184}, 187~\cite{187}, 188~\cite{188}, 189~\cite{189}, 190~\cite{190}, 191~\cite{191}, 192~\cite{192}, 199~\cite{199}, 201~\cite{201}, 209~\cite{209}, 214~\cite{214}, 217~\cite{217}, 218~\cite{218}, 220~\cite{220}, 222~\cite{222}, 223~\cite{223}, 233~\cite{233}, 235~\cite{235}, 238~\cite{238}, 240~\cite{240}, 241~\cite{241}, 245~\cite{245}, 246~\cite{246}, 247~\cite{247}, 248~\cite{248}, 254~\cite{254}, 255~\cite{255}, 256~\cite{256}, 260~\cite{260}, 261~\cite{261}, 264~\cite{264}, 265~\cite{265}, 270~\cite{270}, 275~\cite{275}, 279~\cite{279}, 280~\cite{280}, 281~\cite{281}, 282~\cite{282}, 285~\cite{285}, 288~\cite{288}, and 290~\cite{290}.
\\
\noindent
{\bf Note Added}\\
The final part of the main text has been added because a reviewer at a certain journal expressed interest in the true ground state of ``\emph{hydride superconductivity}''. As I was not given an opportunity to respond, I have included it here.
\\
\\
\\
\noindent
{\small
{\bf Data availability}\\
No measured data or code has been generated in the article.
\\
{\bf Competing interests}\\
The author declares no competing interests.
}
\urlstyle{same}

\end{document}